\documentclass[%
reprint,
showpacs,
 amsmath,amssymb,
 aps,
]{revtex4-1}

\usepackage{graphicx}

\usepackage{subfig}

\begin{document}

\title{From continuous to discontinuous transitions in social diffusion}

\author{P. Tuz\'on\,$^{1}$, J. Fern\'andez-Gracia\,$^{2}$ and V.M. Egu\'iluz\,$^{2}$}
\affiliation{$^{1}$Departament de Did\`actica de les Ci\`encies Experimentals i Socials Facultat de Magisteri, Universitat de Val\`encia, 46022 Val\`encia, Spain \\
$^{2}$Instituto de F\'isica Interdisciplinar y Sistemas Complejos IFISC (CSIC - UIB), E-07122 Palma de Mallorca, Spain}

\date{\today}

\begin{abstract}
Models of social diffusion reflect processes of how new products, ideas or behaviors are adopted in a population. These models typically lead to a continuous or a discontinuous phase transition of the number of adopters as a function of a control parameter. We explore a simple model of social adoption where the agents can be in two states, either adopters or non-adopters, and can switch between these two states interacting with other agents through a network. The probability of an agent to switch from non-adopter to adopter depends on the number of adopters in her network neighborhood, the adoption threshold $T$ and the adoption coefficient $a$, two parameters defining a Hill function. In contrast, the transition from adopter to non-adopter is spontaneous at a certain rate $\mu$. In a mean-field approach, we derive the governing ordinary differential equations and show that the nature of the transition between the global non-adoption and global adoption regimes depends mostly on the balance between the probability to adopt with one and two adopters. The transition changes from continuous, via a transcritical bifurcation, to discontinuous, via a combination of a saddle-node and a transcritical bifurcation, through a supercritical pitchfork bifurcation. We characterize the full parameter space. Finally, we compare our analytical results with Montecarlo simulations on annealed and quenched degree regular networks, showing a better agreement for the annealed case. Our results show how a simple model is able to capture two seemingly very different types of transitions, i.e., continuous and discontinuous and thus unifies underlying dynamics for different systems. Furthermore the form of the adoption probability used here is based on empirical measurements.
\end{abstract}

\pacs{adoption, phase transition, mean-field, social contagion, spreading}

\maketitle

\section{Introduction}

Spreading processes are ubiquitous in nature: from the contagion of diseases \citep{Anderson1991}, herd behaviour in animals \citep{Sumpter2008}, the diffusion of innovations \citep{Rogers2010}, rumour spreading \citep{Daley1965}, the evolution of social movements \citep{Gonzalez-Bailon2013}, the propagation of hashtags in Twitter \citep{Alvarez_2015}, etc. All these processes share similar dynamics; in a population of initially neutral (disease-free, ignorants of some information, etc) agents (humans, animals or even bots), some of them start carrying some information, pathogen, or behavior, i.e. they adopt this innovation. Through a transmission process they can pass it on to other agents, starting in this way the process of adoption diffusion.

The diffusion of adoption has been extensively studied and modeled in several fields including Biology, Physics and Social Sciences \citep{Goel2012,LopezPintado2008,RevModPhys.81.591,RevModPhys.87.925}. In general, new adopters have been in contact with one or several adopters, with two main mechanisms: in disease-like models \citep{Kermack1927,Weiss1971}, adoption takes place with an adoption probability per contact with an adopter which is constant irrespective of the number of adopters; in threshold-like models \citep{LopezPintado2008,Kermack1927,Weiss1971,Granovetter1978}, adoption happens only after a critical number of adopters has been reached. There are also models of ``generalized contagion'' \citep{Dodds2004}, where both disease-like and threshold behaviors are special cases.

However, while the models describe individual adoption probabilities, most of the related empirical research was based on
aggregated data, typically cumulative adoption curves \citep{Bass1969,Young2009}. Recent studies have focused on
individuals' behavior, where the number of adopters accessed by each individual can be measured
\citep{Milgram1969,Dasgupta2008,Romero2011,Gallup2012}. These measurements
have a direct connection with the form of the adoption probability. In this paper we explore the probability function obtained by \citep{Milgram1969} from a social experiment.
They analyzed the correlation
between the size of a group looking at the same point in the street and the number
of passerbies that joined the behavior of looking at that point. The results of the
experiment can be fitted with a Hill function for the probability of adoption \citep{Gallup2012}. We will show that the shape of the adoption probability  leads to two different behaviors depending on
the parameter values: either a continuous or a discontinuous phase transition. This provides a simple model that describes both regimes within the same framework, depending only on two parameters; with a probability function linked to empirical data.

\section{Results}

An agent that has not adopted yet, adopts with
some probability when interacting with an adopter, which turns her an adopter-maker too. After adoption, the agent is ``recovered'' at a certain rate $\mu$
and becomes again a potential adopter. Here, we study the consequences of the probability of adoption. The transition from adopter to non-adopter is assumed to occur at some constant rate $\mu$.

\begin{figure*}[h]
\begin{center}
\renewcommand{\thesubfigure}{A}
\subfloat[\label{a}]{\includegraphics[width=0.45\textwidth]{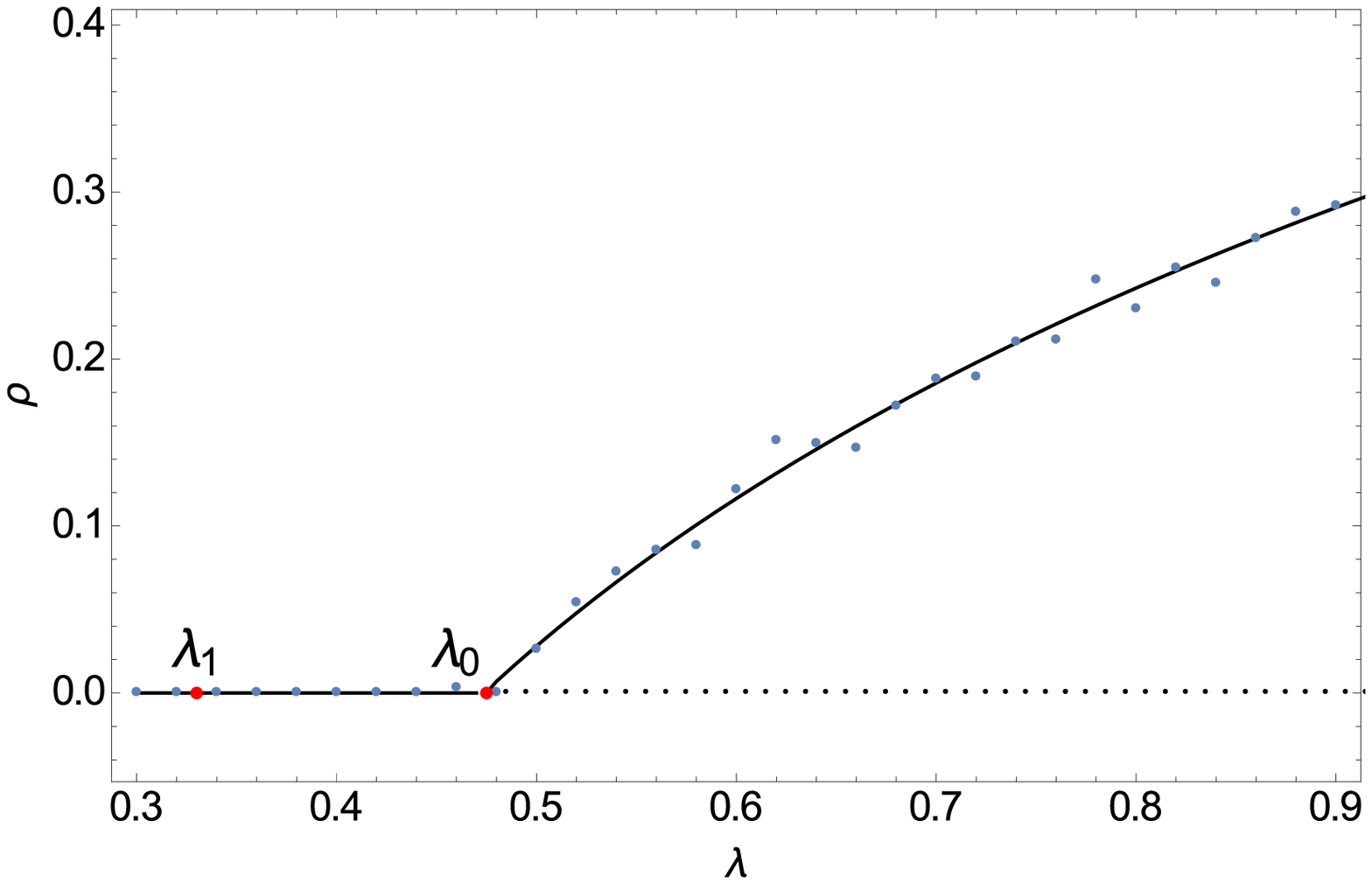}}
\renewcommand{\thesubfigure}{B}
\subfloat[\label{b}]{\includegraphics[width=0.45\textwidth]{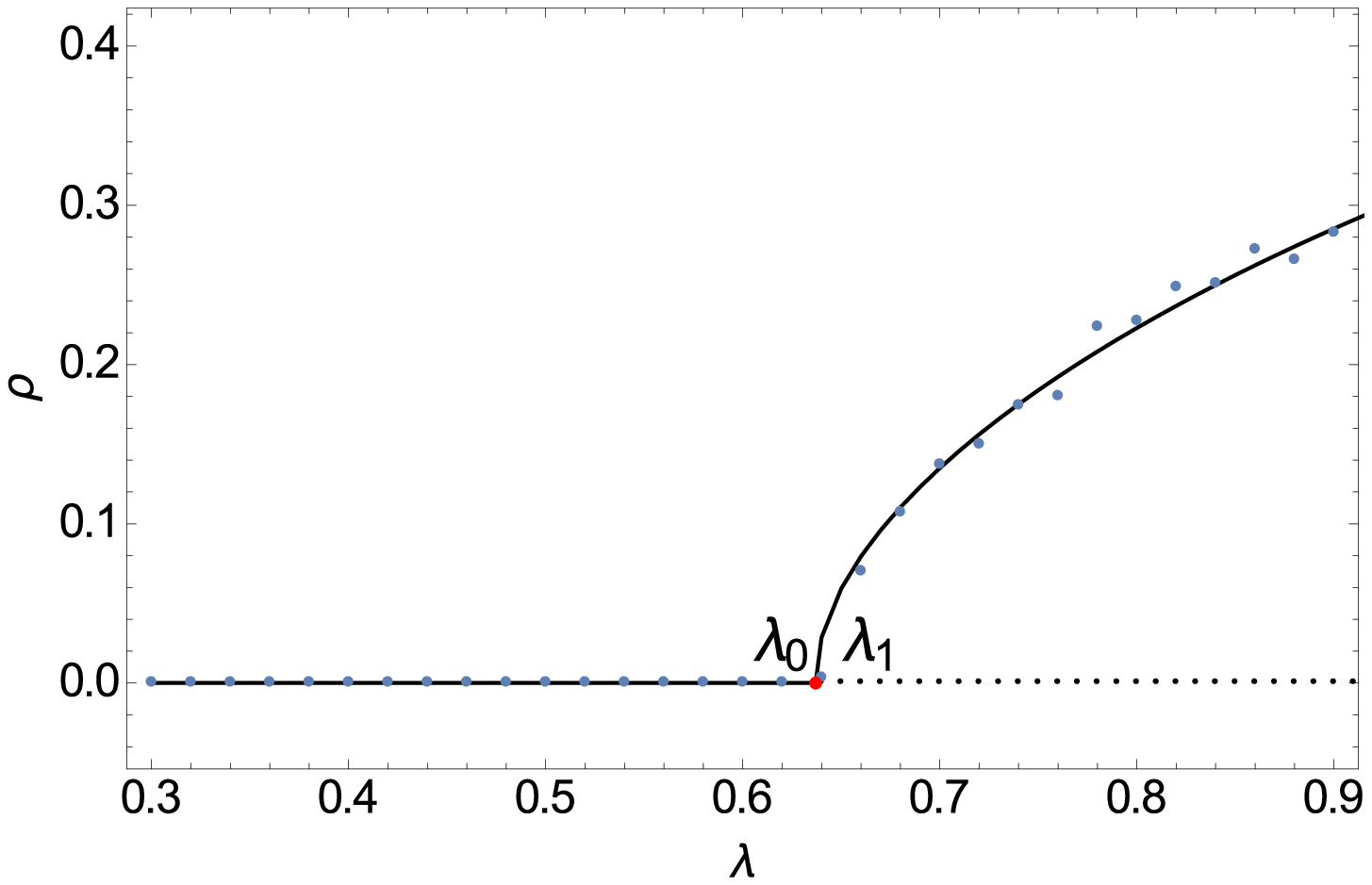}}\\
\renewcommand{\thesubfigure}{C}
\subfloat[\label{c}]{\includegraphics[width=0.45\textwidth]{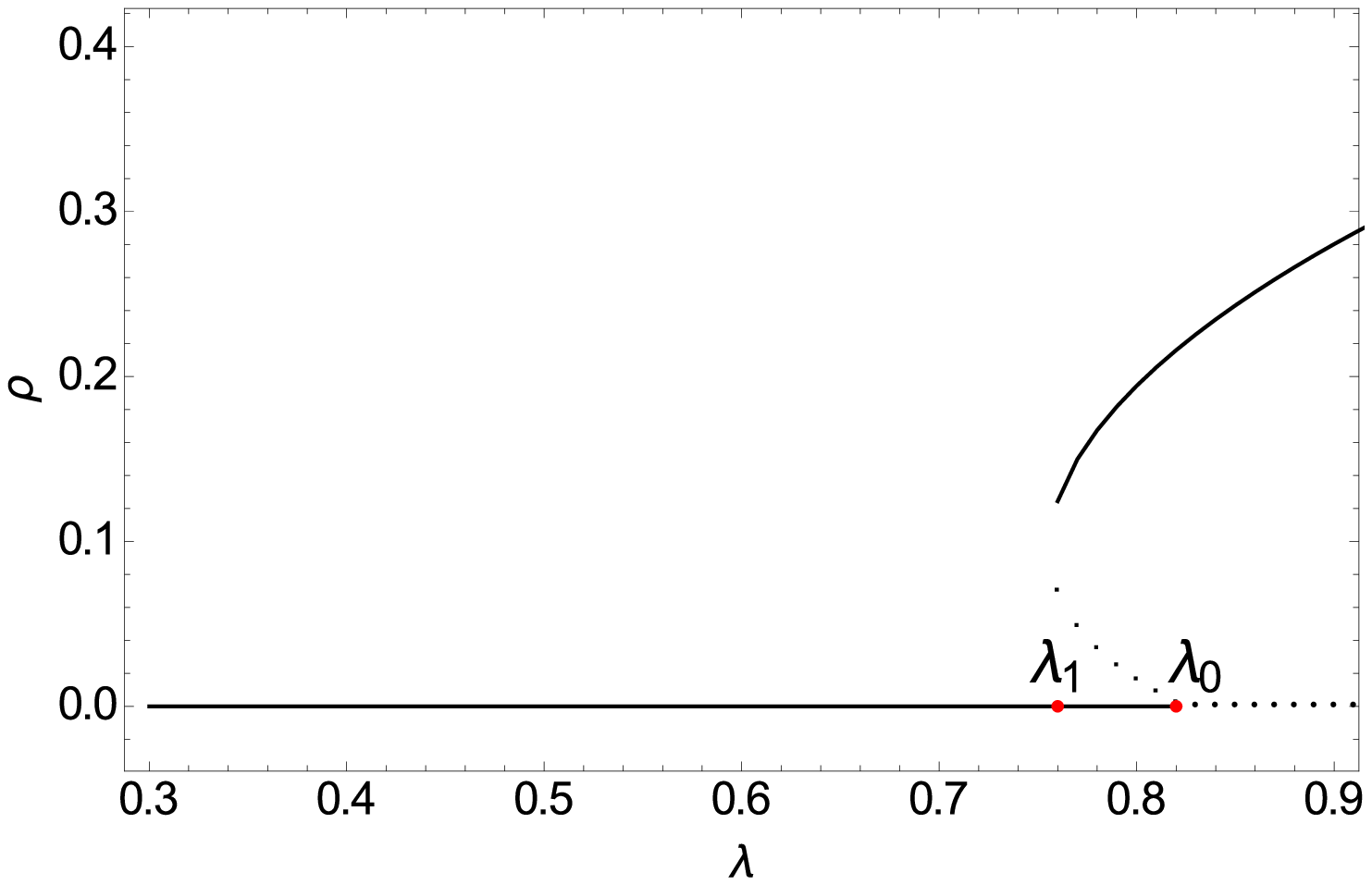}}
\renewcommand{\thesubfigure}{D}
\subfloat[\label{d}]{\includegraphics[width=0.45\textwidth]{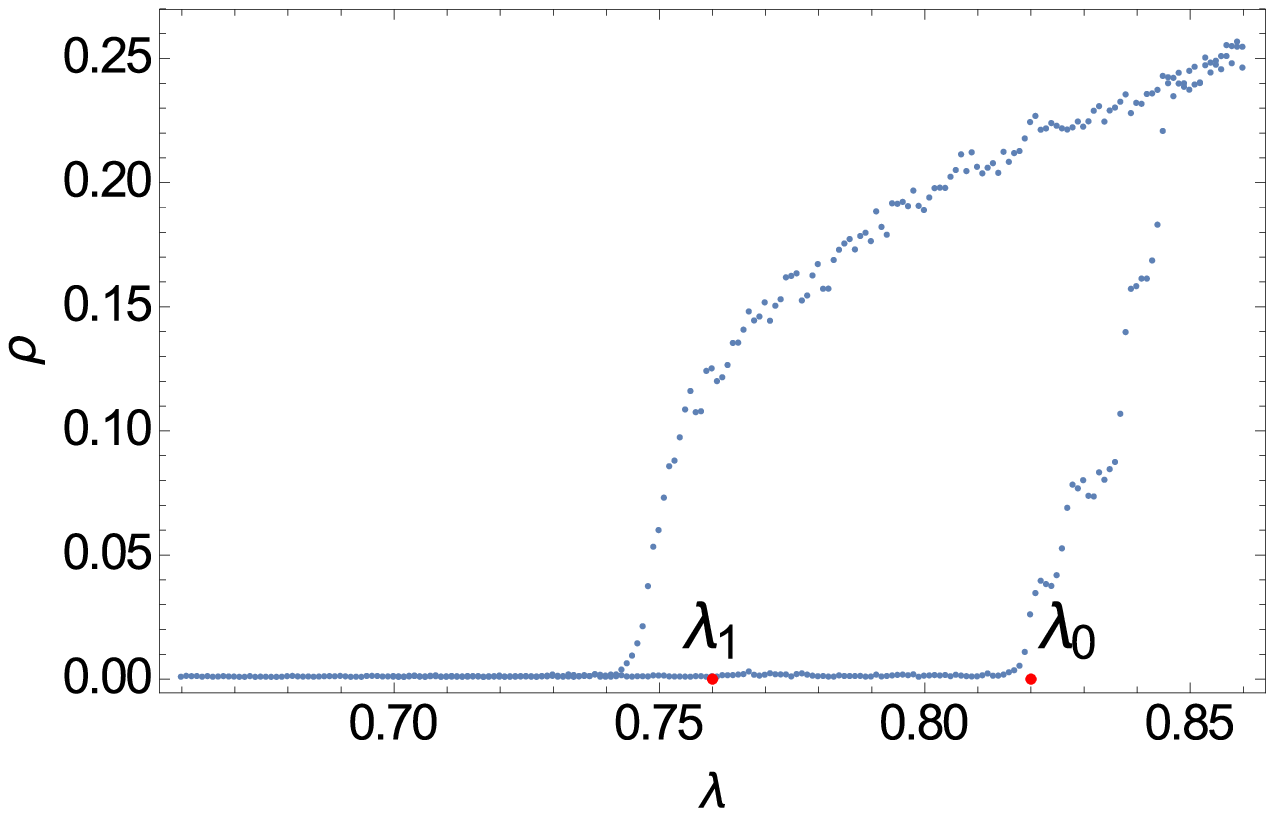}}
\caption{Complete solutions of Eq.~(\ref{eq}) are shown in black for $T=3$, $k=10$ and $a=1.2,1.53, 1.8$ (panels \ref{a}, \ref{b} and \ref{c} respectively). Continuous lines represent stable solutions. Note that when $\lambda_0$ intersects the upper branch, the transition is continuous (\ref{a}). When $\lambda_0$ intersects the lower branch (\ref{c}), two stable solutions coexist in the region $\lambda_1<\lambda<\lambda_0$, 0 and $\rho^{up}$, and the transition is discontinuous. Simulations of the microscopic model are shown in blue points in panels \ref{a} and \ref{b}. For panel \ref{c} the simulation is shown in panel \ref{d}, that amplifies the region $\lambda_1-\lambda_0$, showing the histeresis of the system. Panel \ref{b} illustrates the case when $\lambda_0=\lambda_1$. \label{diagram}}
\end{center}
\end{figure*}

In the standard SIS (susceptible-infected-susceptible) model \citep{Anderson1991}, the adoption probability (from susceptible to infected, S $\to$ I) $\beta$ is constant for each interaction with an adopter. In general, the adoption probability can be a general function of the number of adopted neighbors, $n$ :
\begin{eqnarray}\label{pcomp}
 P(n) = \lambda' f(n) \; .
\end{eqnarray}
In this contribution we will consider the function proposed by Ref.~\citep{Gallup2012}
\begin{eqnarray}\label{Gallup_eq}
f(n) = \frac{n^a}{T^a + n^a} \; ,
\end{eqnarray}
where $\lambda'$ is persuasion capacity (similar to
$\beta=\lambda'$ for $T=0$ and $a=1$), $a$ is the adoption coefficient (or Hill coefficient) and controls how fast/slow
this probability increases with $n$ and $T$ is the adoption threshold and fixes the number of adopters needed to reach half the persuasion limit. $\lambda$, $T$ and $a$ are real positive numbers. This type of function is known as Hill function and has been used in models of population growth and decline \citep{Basios2016a,Gonze2013,santillan2008}.
The evolution of such a system in an annealed degree regular network (a network where all the nodes have the same number of neighbors or degree $k$ but where they are chosen randomly in the population at each interaction)
is determined by
\begin{eqnarray}
 \frac{d\rho}{dt'}&=&-\mu \rho + (1-\rho)  A  ,
\end{eqnarray}
where $\rho$ is the density of adopters and $A$ is the probability of adoption given the density $\rho$ and is given by
\begin{eqnarray}
A = \sum_{n=0}^k P(n) \left( \begin{array}
{c} k \\ n \end{array}\right)\rho^n
(1-\rho)^{k-n} \; .
\end{eqnarray}
The number of infected neighbors is assumed to be binomially distributed with a success probability equal to the global density of infected agents.
Without loss of generality we get rid of parameter $\mu$ by changing the timescale and rescaling the persuasion capacity $\lambda'$
\begin{eqnarray}
 \; t=\mu t'\;\\
 \; \lambda=\frac{\lambda'}{\mu}\;,
\end{eqnarray}
which is equivalent to setting $\mu=1$.
The equilibrium solutions for the system are determined by the condition
\begin{eqnarray}\label{eq}
-\rho^* + (1-\rho^*) A^* =0 \; .
\end{eqnarray}
Given a particular value of $a$ and $T$,
there are at most three possible solutions for $\rho^*$ (Figure~\ref{diagram}):
i) $\rho^*=0$, corresponding to the adoption-free regime, ii) $\rho^*=\rho^{up}$, represented by the upper branch and iii) $\rho^*=\rho^{down}$,
the lower branch. 

The stability of the fixed points can be easily checked by linear stability analysis. The solution $\rho^*=0$ changes stability at
\begin{eqnarray}\label{lcorte}
\lambda_0=\frac{1}{kf(1)},
\end{eqnarray}
being stable for $\lambda<\lambda_0$ and unstable otherwise. As can be seen in Figure~\ref{diagram}, if the solution $\rho^*=0$ intersects the upper branch, then that branch is stable and the solution $\rho^*=0$ changes stability via a transcritical bifurcation. Then for $\lambda>\lambda_0$ and for any initial $\rho_0\neq 0$ the system will end up in the fixed point $\rho^{up}$ (Figure \ref{diagram}\ref{a}). If, on the contrary, the solution $\rho^*=0$ intersects the lower branch, this one is unstable and there is a region $\lambda_1<\lambda<\lambda_0$ for which two stable solutions ($\rho^*=0$ and $\rho^{up}$) coexist, separated by an unstable solution $\rho^{down}$ (Figure \ref{diagram}\ref{b}). For $\lambda=\lambda_1$ the two fixed points of opposite stability annihilate through a saddle-node bifurcation, while at $\lambda=\lambda_0$ we still have a transcritical bifurcation. Therefore in that region the final state of the system will be the upper branch solution $\rho^{up}$ if the initial density $\rho_0>\rho^{down}$ and $0$ otherwise and we can observe hysteresis. For $\lambda>\lambda_0$ and for any initial $\rho_0>0$ the system will end at $\rho^{up}$. Note that $\lambda_0$ is only the critical point for continuous transitions, while for discontinuous ones would be $\lambda_1$.
The sign of the derivative of the $\rho^*$ function at the intersection of $\rho*=0$ and the other branches determines the type of transition.
If the derivative is positive ($\rho^*=0$ intersects $\rho^{up}$), the transition is continuous, while if it is negative ($\rho^*=0$ intersects $\rho^{down}$), the transition
is discontinuous ((\ref{cont}) and (\ref{disc}) respectively).
\begin{subequations}\label{conditions}
\begin{eqnarray}
 \; \left.\frac{d\rho^*}{d\lambda}\right|_{\lambda_0}>0 \; \Longrightarrow \;  f(2) < \frac{2k}{k-1} f(1) \; \label{cont} \\
 \; \left.\frac{d\rho^*}{d\lambda}\right|_{\lambda_0}<0 \; \Longrightarrow \; f(2) > \frac{2k}{k-1} f(1) \; . \label{disc}
\end{eqnarray}
\end{subequations}

For the particular case when $f(2)=\frac{2k}{k-1} f(1)$ both $\lambda_0$ and $\lambda_1$ coincide. For this condition one can show, by approximating Eq.~\ref{eq} to third order in $\rho*$, that the bifurcation diagram is that one of a supercritical pitchfork bifurcation, i.e., the equation is equivalent to $\dot{x}=rx-x^3$ (Figure \ref{diagram}\ref{c}). In this case, the final fate of the system is similar to the continous case. For $\lambda<\lambda_0$ there is no global adoption and the system ends at $\rho^*=0$, while for $\lambda>\lambda_0$ any initial condition $\rho_0\ne0$ will bring the system to $\rho^{up}$.

Simulations using a microscopic model are also included in the plots of Figure \ref{diagram}. This microscopic model simulates an SIS dynamics in a degree regular network of $k=10$ that changes at each timestep. From one step to another, an agent is selected; if
it is an adopter it recovers with probability $\mu$, if not, it adopts
with probability $P(n)$, where $n$ is the number of adopters among $k$ randomly chosen agents. There is an
initial seed of infected agents which we fix to $1\%$ of the total population.

In pannels \ref{a} and \ref{b} of Figure \ref{diagram} results of the simulations are shown in blue dots over the analytical solution. For panel \ref{c}, simulations are shown in panel \ref{d}. As can be seen, the system exhibits hysteresis in the region $\lambda_1 < \lambda < \lambda_0$, where there is bistability. There system ends at $\rho^{up}$ or $\rho^{down}$ depending on the initial condition.

Fig.~\ref{diagram} also illustrates the two different kinds of transitions. The density of adopters stays at zero until
a critical value of $\lambda$, where the system goes to $\rho^{up}$ by either a continuous transition
or a discontinuous transition. As can be observed, provided a value for $T$, the
size of the jump increases with $a$. For values of $a \sim 1$ the system resembles
the epidemic-like models while for values $a>1$ the transition
is threshold-like.

\begin{figure*}[h]
\setcounter{subfigure}{0}
\begin{center}
\renewcommand{\thesubfigure}{A}
\subfloat[\label{3a}]{\includegraphics[width=0.45\textwidth]{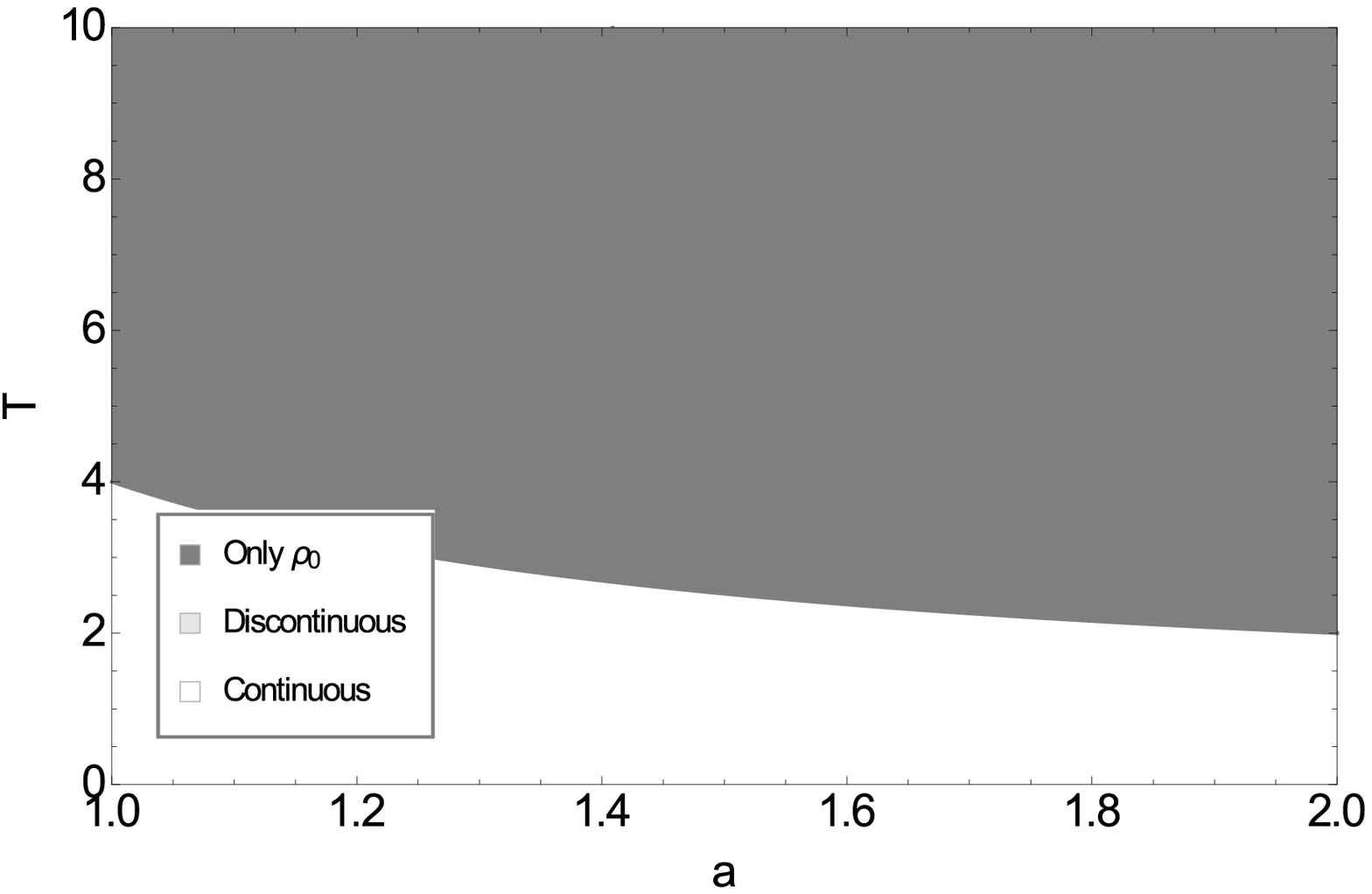}}
\renewcommand{\thesubfigure}{B}
\subfloat[\label{3b}]{\includegraphics[width=0.45\textwidth]{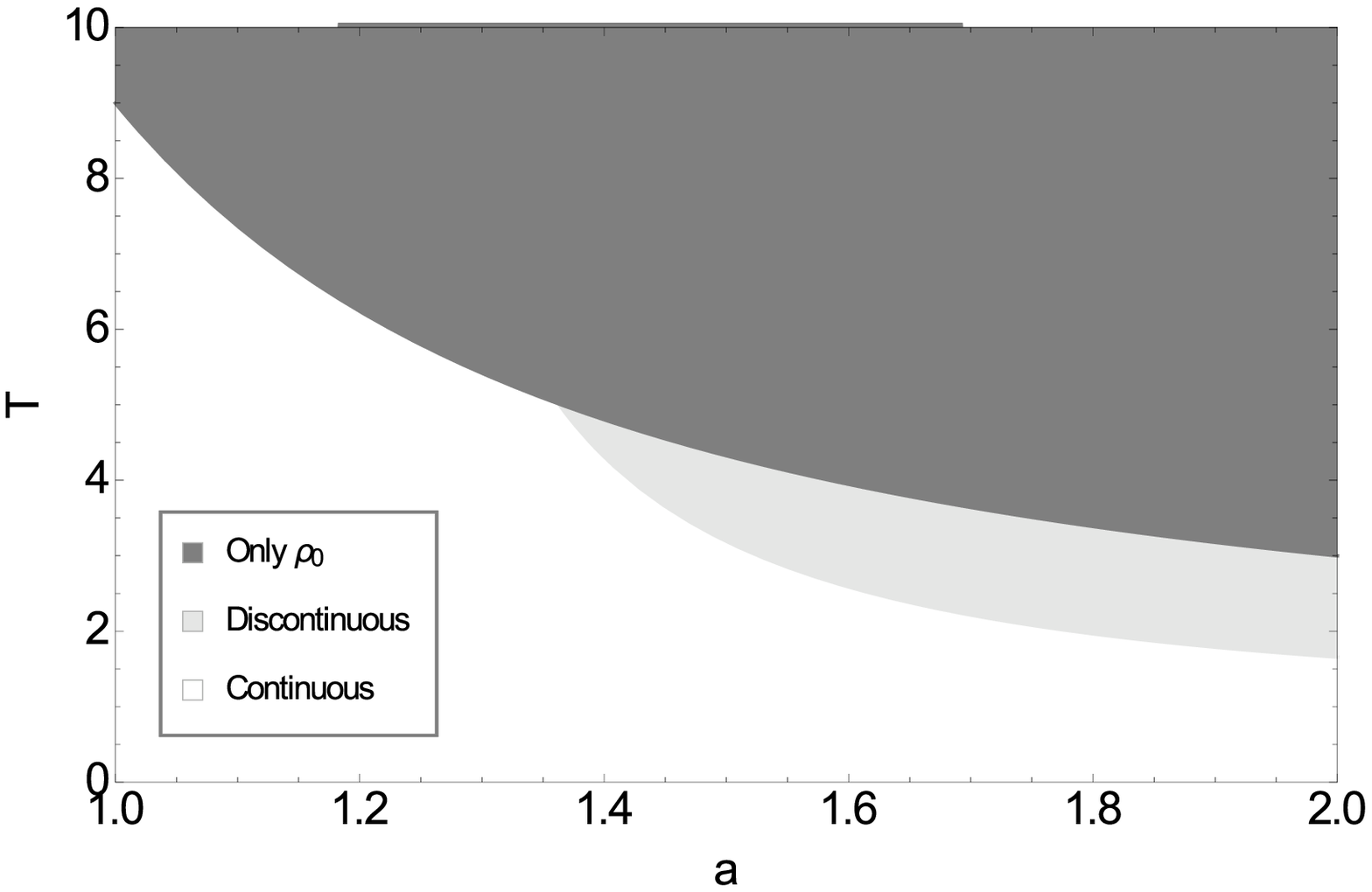}}\\
\renewcommand{\thesubfigure}{C}
\subfloat[\label{3c}]{\includegraphics[width=0.45\textwidth]{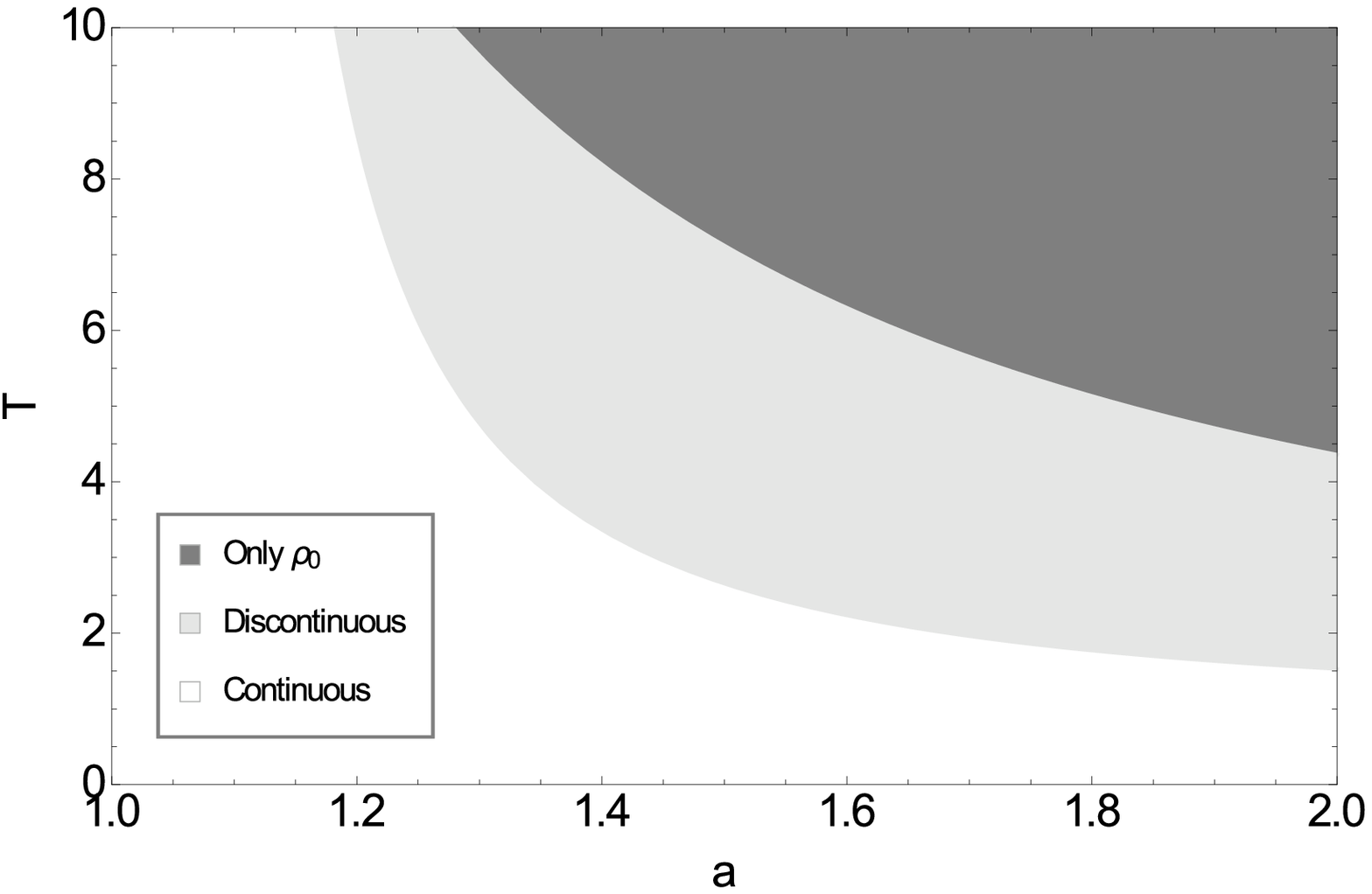}}
\caption{Parameter space for a regular random network with $k=5,10,20$ (panels \ref{3a}, \ref{3b} and \ref{3c} respectively). The white area is for contininuous transitions while the light gray area is for discontinous transitions. Both areas are separated by the curve given by equation \eqref{conditions2}, corresponding to a supercritical pitchfor bifurcation diagram. In the dark gray area only the solution $\rho^*=0$ exists, i.e., there is not global adoption.\label{discontR}}
\end{center}
\end{figure*}

For the case of our choice of $f(n)$ (Eq.~\ref{Gallup_eq}) the conditions in Eqs.~\ref{conditions} give bounds for the parameters region for which the transition is of one regime or the other:
\begin{subequations}\label{conditions2}
\begin{eqnarray}
\text{Cont.: } && T <  \left(  \frac{2^a (k+1)}{2^a(k-1)-2k}  \right)^{\frac{1}{a}} \; \label{cont2} \\
\text{Disc.: } && T > \left(  \frac{2^a (k+1)}{2^a(k-1)-2k}  \right)^{\frac{1}{a}} \; . \label{disc2}
\end{eqnarray}
\end{subequations}
Fig.~\ref{discontR} shows this
parameters space
 for $k=5,10,20$. The white region represents the parameters combination for a continuous transition while
the light gray region corresponds to a discontinuous transition. The dark gray region is the condition that
$\lambda_0 \leq1$
on Eq. (\ref{lcorte}), that is, that the value where both curves meet is in the range $\lambda \le 1$,
\begin{eqnarray}\label{condition0}
T< (k-1)^{\frac{1}{a}} \; .
\end{eqnarray}
This constraint implies that the in dark gray region in the plot there is only one possible solution, $\rho*=0$.

Both conditions together, Eq.~\ref{conditions2} and \ref{condition0}, predict the values of the parameters for which the model shows one type of transition or another, or none. For example, in panel \ref{3b} of Figure~\ref{discontR}, a continuous transition is allowed
for all values of $a \in [1,2]$ and some values of $T \in [0,10]$, while the discontinuous transition is only possible for values of $a$ higher than 1.25 and values of $T$ higher than 1.5. As can be seen in Figure~\ref{discontR}, for small values of $k$, there are only continuous transitions, while for higher values of $k$, also discontinuous transitions are allowed. Besides, the higher the value of $k$, the more paramater space allows for $\rho \neq 0$ solutions.

\begin{figure*}[h]
\begin{center}
\includegraphics[width=.6\textwidth]{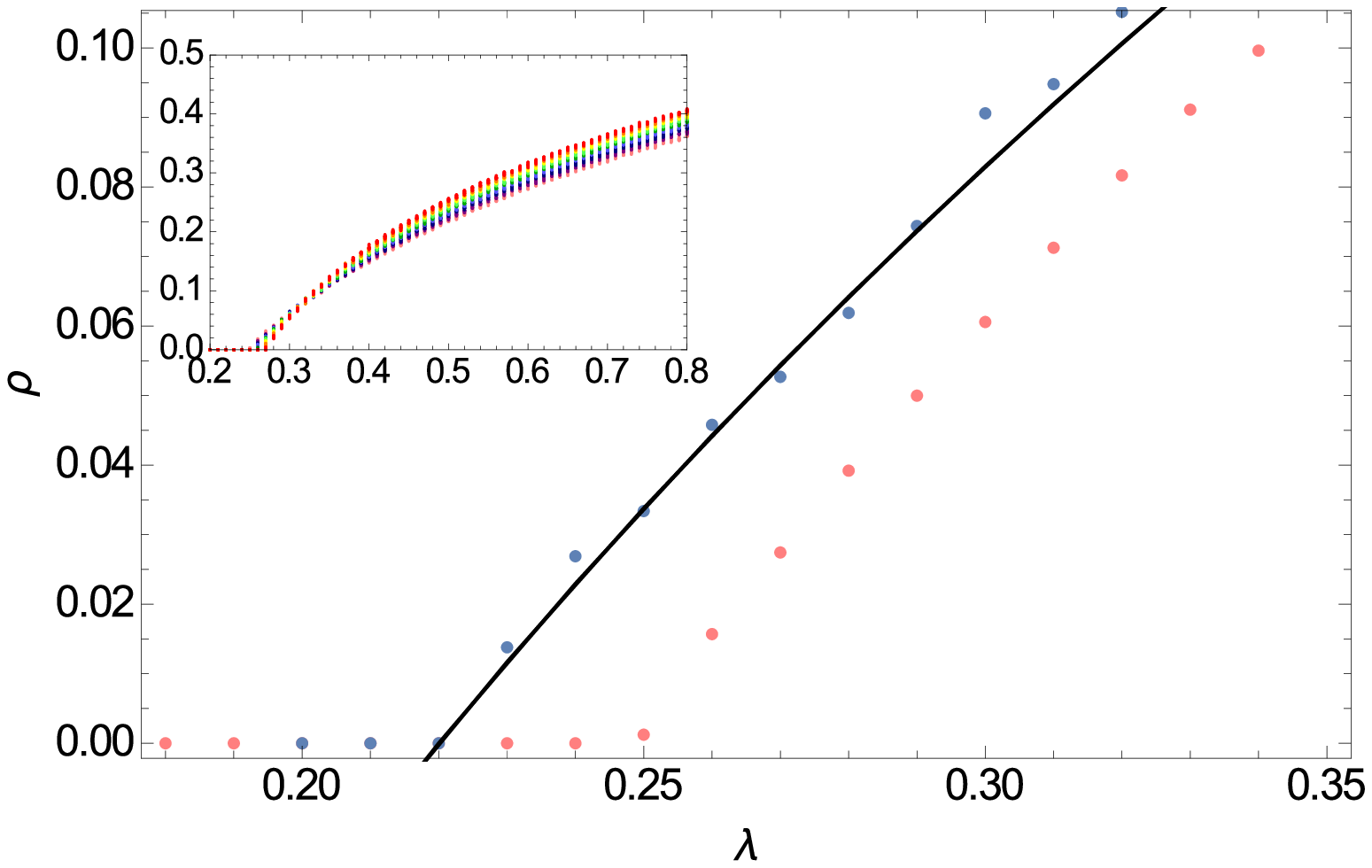} 
\includegraphics[width=.6\textwidth]{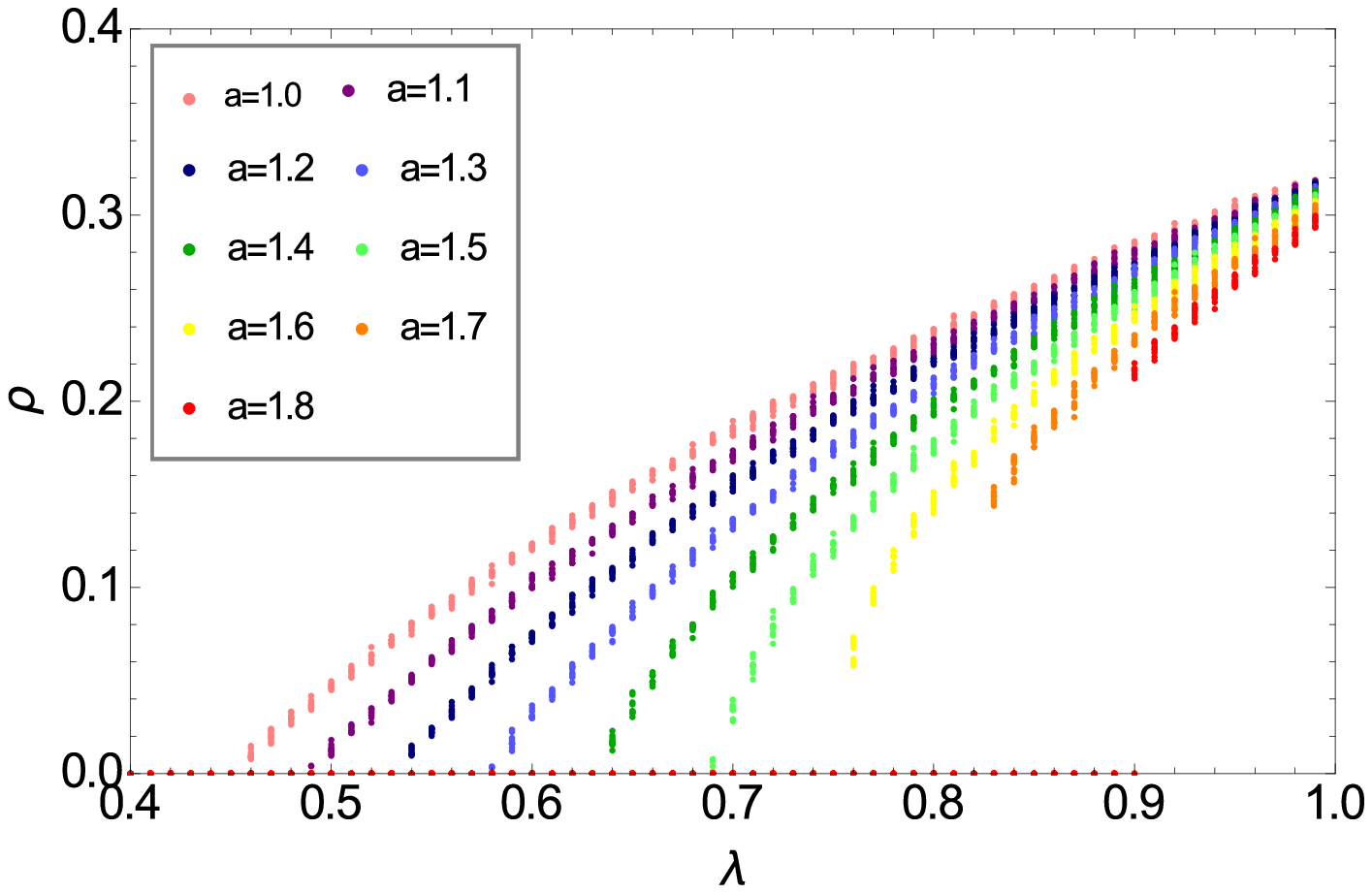}
\caption{Simulations of the microscopic model on a degree regular random network with degree $k=10$. Individuals might adopt with probability $P(n)$. Upper panel shows the results for $T=1.2$ and lower panel for $T=3$ for different values of $a$. For $T=1.2$ the transitions are continuous for any $a$ (inset, same color code as lower panel). The upper panel shows the region of the critical point for the simulations of the microscopic model the quenched network (pink), the simulations on the annealed network (blue) and the exact solution (black line) of the equation for $a=1.0$ respectively. For $T=3$ there are continuous or discontinous transitions depending on the value of $a$.\label{simR}}
\end{center}
\end{figure*}

Finally, we perform simulations to characterize numerically the behavior of the system using a similar microscopic model on quenched regular random network. Again, at each timestep an agent is selected, if
she is an adopter it recovers with probability $\mu$, if not, she adopts
with probability $P(n)$, where now $n$ refers to the number of adopters in her network neighborhood, which is now fixed. There is an
initial seed of infected agents equal to $1\%$ of the total population.
The long term values of the fraction of adopters $\rho_{\infty}$ are shown in Figure~\ref{simR} for 10 realizations and different values
of $a$ for $T=1.2,3$. The realizations are not averaged to show the low dispersion (inset of upper panel in Figure~\ref{simR} and lower panel of Figure~\ref{simR}).

As Fig.~\ref{discontR} indicates for $T=1.2$ and $k=10$, the system exhibits always a continuous transition no matter the values of $a\in [1,2]$ (inset of the upper panel).
For $T=3$ and $k=10$, for values of $a$ higher than 1.5 the transition is discontinuous, as shown in Figure~\ref{discontR}. The upper panel of Figure \ref{simR} zooms in the region of the critical point for the case of $a=1.0$. It shows the simulations of the microscopic model on a quenched degree regular random network (pink), on an annealed degree regular random network (blue) and the exact solution of the equation (black). As can be seen, there is a small discrepancy for the model on the quenched version of the network. This is because when the topology is fixed correlations appear and in particular the approximation that the infected agents are binomially distributed among the neighbors with a success probability equal to the global fraction of infected agents breaks down. As in the cases presented above, the simulations won the annealed network and the exact solution agree. For both microscopic models, the type of transition is predicted by the parameters space represented in Figure \ref{discontR}.

\section{Conclusions}

We have analyzed a model of social contagion (SIS-like) on degree regular random networks with an adoption probability measured in empirical data in \citep{Gallup2012} that interpolates between the cases of epidemic-like spreading and threshold-like dynamics. We show that this simple model displays both continuous and discontinuous transitions from a disease-free state to an endemic state. We find the values of the parameters that separate this transitions and the critical persuasion capacities $\lambda$ by applying standard linear stability and bifurcation theory tools.

The simplicity of the model studied here allows for relaxing some of the assumptions considered here. For example, the stability condition given by Eq.~\ref{lcorte} resembles the structure of the critical point in the SIS model in uncorrelated random networks with arbitrary degree distributions. Following this similarity, we conjecture that the solution of our model in complex networks will be given by  $\lambda_0= <k>/<k^2> f(1)$. Thus degree heterogeneity will lead to the vanishing threshold unless $f(1) \to 0$ as $N\to infinity$. This can be achieved for example by considering that $T= c k_max$. Alternatively, an interesting variation is to consider that the adoption probability depends not on the absolute number of adopters but on the fraction of them. Besides, heterogeneity can emerge not only at the degree level, but also in the distributions of the adoption threshold $T$ and adoption coefficient $a$ and furthermore they can be correlated with the degree of the nodes. How heterogeneity affects the nature of the transition needs to be explored in detail. Another possible line of research is adding non-Markovianity to the dynamics, for example by letting the adoption probability depend not only on the state of the neighboring agents, but also on some internal time which takes into account when an agent tries to convince another one for adopting the innovation.

Our results highlight that not only the structure of the interaction network neither the dynamics alone are responsible of the type of transition that the system displays. Furthermore this simplified framework is able to capture this seemingly disparate types of transition, which are usually taken as a signature of different dynamics. Furthermore the choice of the adoption probability curve is based on empirical measurements from \citep{Gallup2012}, which highlights the relevance of our results for realistic modeling of social phenomena.

\end{document}